\documentclass[%
 aip, reprint,
 amsmath,amssymb,
nofootinbib]{revtex4-2}

\usepackage{xcolor}

\usepackage{hyperref}
\hypersetup{colorlinks=true, citecolor=blue, urlcolor=blue, linkcolor=blue}

\usepackage{dcolumn,amsmath}
\usepackage[T2A]{fontenc}
\usepackage[utf8]{inputenc}
\usepackage{graphicx}
\graphicspath{ {./Images/} }
\usepackage{bm}
\usepackage{xfrac}
\usepackage{indentfirst} 
\usepackage{physics}

\usepackage{xcolor}

\usepackage{color}

\begin{document}

\title{Refined theoretical values of field and mass isotope shifts in thallium to extract charge radii of Tl isotopes}

\date{11.01.2023}

\author{Gleb Penyazkov}
\email{glebpenyazkov@gmail.com}
\affiliation{Petersburg Nuclear Physics Institute named by B.P.\ Konstantinov of National Research Center ``Kurchatov Institute'' (NRC ``Kurchatov Institute'' - PNPI), 1 Orlova roscha mcr., Gatchina, 188300 Leningrad region, Russia}
\affiliation{Saint Petersburg State University, 7/9
Universitetskaya Naberezhnaya, St. Petersburg, 199034 Russia}

\author{Sergey D. Prosnyak} 
\affiliation{Petersburg Nuclear Physics Institute named by B.P.\ Konstantinov of National Research Center ``Kurchatov Institute'' (NRC ``Kurchatov Institute'' - PNPI), 1 Orlova roscha mcr., Gatchina, 188300 Leningrad region, Russia}
\affiliation{Saint Petersburg State University, 7/9
Universitetskaya Naberezhnaya, St. Petersburg, 199034 Russia}

\author{Anatoly E. Barzakh} 
\affiliation{Petersburg Nuclear Physics Institute named by B.P.\ Konstantinov of National Research Center ``Kurchatov Institute'' (NRC ``Kurchatov Institute'' - PNPI), 1 Orlova roscha mcr., Gatchina, 188300 Leningrad region, Russia}

\author{Leonid V. Skripnikov}
\homepage{http://www.qchem.pnpi.spb.ru}
\email{skripnikov\_lv@pnpi.nrcki.ru, leonidos239@gmail.com}
\affiliation{Petersburg Nuclear Physics Institute named by B.P.\ Konstantinov of National Research Center ``Kurchatov Institute'' (NRC ``Kurchatov Institute'' - PNPI), 1 Orlova roscha mcr., Gatchina, 188300 Leningrad region, Russia}
\affiliation{Saint Petersburg State University, 7/9
Universitetskaya Naberezhnaya, St. Petersburg, 199034 Russia}

\begin{abstract}    

Electronic factors for the field and mass isotope shifts in the  $6p ^{2}P_{3/2} \to 7s ^{2}S_{1/2}$ (535 nm), $6p ^{2}P_{1/2} \to 6d ^{2}D_{3/2}$ (277 nm) and  $6p ^{2}P_{1/2} \to 7s ^{2}S_{1/2}$ (378~nm) transitions in the neutral thallium were calculated within the high-order relativistic coupled cluster approach. These factors were used to reinterpret previous experimental isotope shift measurements in terms of charge radii of a wide range of Tl isotopes. Good agreement between theoretical and experimental King-plot parameters was found for the $6p ^{2}P_{3/2} \to 7s ^{2}S_{1/2}$ and $6p ^{2}P_{1/2} \to 6d ^{2}D_{3/2}$ transitions. It was shown that the value of the specific mass shift factor for the $6p ^{2}P_{3/2} \to 7s ^{2}S_{1/2}$ transition is not negligible compared to the value of normal mass shift in contrary to what had been suggested previously. Theoretical uncertainties in mean square charge radii were estimated. They were substantially reduced compared to the previously ascribed ones and amounted to less than 2.6\%. The achieved accuracy paves the way to a more reliable comparison of the charge radii trends in the lead region.

\end{abstract}

\maketitle

\section{Introduction}
It is widely acknowledged that nuclear charge radii provide sensitive tests of different aspects of nuclear structure~\cite{Yang2022_exotic}. High precision charge radii of unstable nuclei prove to be a benchmark for state-of-the-art nuclear models~\cite{Garcia2016,deGroote2020,Koszorus2021,Goodacre2021}. Apart from this, the precise charge radii of some isotopes can also be used to constrain the parameters of the nuclear matter~\cite{Pineda2021}. Changes in mean-square (ms) charge radii can be extracted from the isotope shift (IS) of the atomic transitions. Isotope shift is represented as a sum of two contributions: field shift (FS), which is associated with a change in the nucleus charge distribution, and the mass shift (MS), connected with the nuclear recoil effect~\cite{Yang2022_exotic}. The mass shift contribution can be further separated into the normal mass shift (NMS) and specific mass shift (SMS). Each contribution can be factorized into the electronic and nuclear parts~\cite{Yang2022_exotic}. 
Corresponding electronic multipliers are denoted below as FS and MS (NMS and SMS) factors. 
To extract the change in the ms charge radius from the experimental IS values, one needs to know the electronic factors. Progress in the experimental technique enables one to obtain IS data with an accuracy ranging between 2\% and 0.05\% (see, for example, Refs.~\onlinecite{Koszorus2021,Han2022,Barzakh2021,Garcia2016,deGroote2020}). However, until recently electronic factors were determined (either by atomic calculations or comparison with the mesoatomic and K X-ray data) without assigning the uncertainty~\cite{Fricke2004NuclearCR}. Empirically estimated theoretical uncertainties (“of about 10\%, or even more in some cases”), proposed by Otten~\cite{Otten1989} for the FS factor, were far larger compared with the achieved experimental precision. The MS factor was estimated, as a rule, qualitatively with an indeterminacy of $\sim$100\% and higher. As a result, the uncertainty stemming from the calculated atomic quantities dominated the overall uncertainty of the studied nuclear observables. Growing requirements for the accuracy of the experimental charge-radii values to catch, in particular, the subtle nuclear effects, for example, the influence of the higher-order nuclear radial moments~\cite{Papoulia2016}, became a challenge to the atomic theory. The development of the advanced electronic-structure calculation methods gives us the tool to obtain reliable theoretic uncertainties and reduce them to the acceptable level (3\% and lower) (see, for example, Refs.~\onlinecite{Kalita2018,Han2022,Sahoo_2020,Gustafsson2020}).

Isotope shifts in the neutral thallium atoms (A = 179–208) were extensively studied experimentally. Three different atomic transitions were considered: $6p ^{2}P_{\sfrac{3}{2}} \to 7s ^{2}S_{\sfrac{1}{2}}$ (535 nm)~\cite{PhysRevLett.55.1559,Hull:61,Odintsov1960,Hermann1993,PhysRevC.36.2560,GOORVITCH19671,Davis:66,Menges1992,PhysRev.188.1897,Buchinger1992,SCHUESSLER1995583}, $6p ^{2}P_{\sfrac{1}{2}} \to 6d ^{2}D_{\sfrac{3}{2}}$ (277 nm)~\cite{Barzakh2013,Hermann1993,Barzakh2017}, and $6p ^{2}P_{\sfrac{1}{2}} \to 7s ^{2}S_{\sfrac{1}{2}}$ (378 nm)~\cite{PhysRevLett.68.1675,Hull:61,Schuler:62,PhysRevA.62.012510,Davis:66,GOORVITCH19671,PhysRev.188.1897}. In the case of 535 and 277 nm lines, the consistency of the corresponding atomic factors was ensured by the King-plot\footnote{It is a two-dimensional plot with the modified experimental IS value for one transition on the horizontal axis and modified experimental IS value for another transition on the vertical axis; the corresponding dependence for different isotopes pairs is expected to be linear, see e.g., Refs. \onlinecite{Yang2022_exotic,Otten1989}.} procedure~\cite{Barzakh2013}, whereas the 378 nm line was treated independently, and it remained unclear whether these data are consistent with those of the other transitions. Correspondingly, IS for $^{208}$Tl, which was measured only for 378-nm line~\cite{PhysRevLett.68.1675}, could not be reliably correlated with the data of other isotopes. Apart from this, the FS factors calculated for the 535 nm transition by different theoretical approaches were ranging between 15.65 and 20.75 GHz/fm$^2$, i.e., there was indeterminacy of $\sim$30\%. It should be stressed that the thallium chain belongs to the lead region, where there is a striking diversity of the patterns of the charge radii isotopic dependency~\cite{Barzakh2021,Marsh2018}. Large indeterminacy of the FS factor hinders the reliable comparison of the thallium isotopic chain with that of adjacent lead or bismuth isotopes.

In the present, paper we refine theoretical values of FS, NMS and SMS factors for $6p ^{2}P_{\sfrac{1}{2}} \to 7s ^{2}S_{\sfrac{1}{2}}$, $6p ^{2}P_{\sfrac{3}{2}} \to 7s ^{2}S_{\sfrac{1}{2}}$ and $6p ^{2}P_{\sfrac{1}{2}} \to 6d ^{2}D_{\sfrac{3}{2}}$ transitions using sophisticated electronic structure methods. For this, we developed a scheme that uses the relativistic coupled cluster theory with the inclusion of high-order cluster amplitudes. An important feature of the scheme is that it allows us to estimate uncertainties in a systematic way. Using calculated parameters and available experimental data on IS, we reexamine the changes in the ms charge radii for Tl isotopes.

\section{Theory and computational details}

The following parametrization of the isotope shift of the transition energy $\Delta \nu^{A', A} = \nu^{A'}-\nu^{A}$ was used:
\begin{equation}
\label{freq}
    \Delta \nu^{A', A} = (k_{\rm NMS}+k_{\rm SMS})(\frac{1}{M^{A'}}-\frac{1}{M^{A}})+F\delta\left\langle r^2 \right\rangle^{A', A}.
\end{equation}
Here $k_{\rm NMS}$, $k_{\rm SMS}$ are normal and specific mass shift constants, $F$ is the field shift constant, $M^{A}$ and $M^{A'}$ are the masses of isotopes with mass numbers $A$ and $A'$, $\delta\left\langle r^2 \right\rangle^{A', A}=\left\langle r^2 \right\rangle^{A'}-\left\langle r^2 \right\rangle^{A}$ is the difference between ms nuclear charge radii of the isotopes. The field shift constant $F$ was defined as $F=\partial \nu/\partial \left\langle r^2 \right\rangle$
at the point of $\sqrt{<r^2>}=5.4759$~fm corresponding to $^{205}$Tl~\cite{ANGELI201369}, where $\nu$ is the electronic transition energy.
The nuclear recoil constants $k_{\rm NMS}$ and $k_{\rm SMS}$ can be calculated using the following relativistic operators~\cite{shabaev1985mass, palmer1987reformulation, shabaev1988nucl, shabaev1994relativistic}:
\begin{equation}
\label{opNMS}
    H_{\rm NMS} = \frac{1}{2M}\sum_i(\vec{p}_i^2-\frac{\alpha Z}{r_i}\left[ \vec{\alpha}_i+\frac{(\vec{\alpha}_i\cdot \vec{r}_i)\vec{r}_i}{r_i^2}\right]\cdot\vec{p}_i),
\end{equation}
\begin{equation}
\label{opSMS}
    H_{\rm SMS} = \frac{1}{2M}\sum_{i \neq k}(\vec{p}_i\cdot\vec{p}_k-\frac{\alpha Z}{r_i}\left[ \vec{\alpha}_i+\frac{(\vec{\alpha}_i\cdot \vec{r}_i)\vec{r}_i}{r_i^2}\right]\cdot\vec{p}_k),
\end{equation}
where $Z$ is the proton number, $\vec{\alpha}_i$ are Dirac matrices corresponding to electron $i$, and $\vec{r}_i$ is the coordinate of $i$-th electron. 
Note that $H_{\rm SMS}$ is a two-electron operator.

In the present work, a new code for the calculation of matrix elements of the operators~(\ref{opNMS}) and~(\ref{opSMS}) over four-component atomic bispinors expressed in terms of the Gaussian-type basis sets was developed. The code was interfaced to highly efficient relativistic electronic structure packages {\sc dirac}~\cite{DIRAC19,Saue:2020}, {\sc mrcc}~\cite{MRCC2020,Kallay:1,Kallay:2} and Exp-T~\cite{EXPT_website,Oleynichenko_EXPT}.

\begin{table}[h]
    \caption{The composition of the basis sets used in calculations. Basis sets are given in the descending order of their quality.}
    \centering
    \begin{tabular}{ll}
    \hline
    \hline
      Basis set   &  Functions\\
      \hline
      Lbas   &  44$s$, 40$p$, 25$d$, 16$f$, 10$g$, 5$h$, 2$i$\\
      MbasExt   &  33$s$, 29$p$, 22$d$, 13$f$, 4$g$, 1$h$\\
      Mbas   &  33$s$, 29$p$, 20$d$, 12$f$, 4$g$, 1$h$\\
      Sbas   &  24$s$, 20$p$, 14$d$, 9$f$, 1$g$\\
      \hline
      \hline
      \end{tabular}
\label{table0}
\end{table}

To calculate the electronic factors in Tl, we implemented the following computational scheme, similar to that used in our studies of other properties of heavy atoms and molecules~\cite{Skripnikov:16b,Prosnyak:2021,Skripnikov:2020e,Prosnyak:2020,Skripnikov:17b}. In the first step, we used the relativistic coupled cluster method with single, double, and perturbative triple cluster amplitudes, CCSD(T), within the Dirac-Coulomb Hamiltonian to obtain a balanced description of both relativistic and electron-correlation effects. In this calculation, all electrons of Tl were included in the correlation treatment. 
In the calculation of FS and NMS factors, the uncontracted Dyall's AE4Z~\cite{Dyall:98,Dyall:06,Dyall:12} basis set augmented by several diffuse functions was used. This basis set is referred to as LBas. Its composition is given in Table~\ref{table0}. In the calculation of the SMS factors, we used the extended Dyall's AE3Z~\cite{Dyall:98,Dyall:06,Dyall:12} basis set. This basis set is referred to as MBasExt in Table~\ref{table0}.
In the next step, several corrections to the values of FS and MS factors obtained in the first step were applied. The first of them is a correction on higher-order correlation effects. For this, we performed a series of coupled cluster calculations in which, in addition to the CCSD(T) model, the CCSDT (the coupled cluster with single, double, and iterative triple amplitudes) as well as the CCSDT(Q) (the coupled cluster with single, double, triple, and perturbative quadruple cluster amplitudes)~\cite{Kallay:6,MRCC2020} methods were used. This series was carried out for 21 valence and outer-core electrons of Tl using the extended Dyall's AE3Z basis set, MBas (see Table~\ref{table0}). For the uncertainty estimation, we carried out a similar calculation within the smaller SBas basis set based on the Dyall's AE2Z one~\cite{Dyall:98,Dyall:06,Dyall:12} (see Table~\ref{table0}).
Next, we estimated the correction due to the increase of the basis set for the SMS-factor calculation as a difference between SMS factor values calculated in the LBas and MBasExt basis sets using all-electron relativistic CCSD (the coupled cluster with single and double amplitudes) approach. 
The effect of the Gaunt inter-electron interaction was estimated as a difference between results obtained within the Dirac-Coulomb-Gaunt Hamiltonian and Dirac-Coulomb Hamiltonian at the self-consistent-field (SCF) level for the $k_{\rm NMS}$ and $k_{\rm SMS}$ factors and the relativistic Fock-Space coupled cluster singles and doubles (FS-CCSD) level for the $F$ factor. In the latter calculation, we used the two-component all-electron Hamiltonian employing the X2C technique within the molecular mean-field approximation~\cite{Sikkema:2009}. Moreover, the following corrections for the $F$ factor were applied. We considered the influence of the remaining part (retardation) of the zero-frequency Breit interaction within the SCF approach, implemented in the {\sc hfd} code~\cite{HFD,Bratzev:77,HFDB}. All calculations described above were performed for the Gaussian nuclear charge distribution model~\cite{Visser:87}. To take into account more realistic charge distribution, we added a correction calculated as the difference between the $F$ factor values obtained within the Fermi and Gaussian charge distribution models at the Dirac-Hartree-Fock level~\cite{HFD,Bratzev:77,HFDB}.

\begin{table*}
\caption{Calculated values of the field shift constant $F$ (in GHz/fm$^2$) and their uncertainties for the $6p ^{2}P_{\sfrac{3}{2}} \to 7s ^{2}S_{\sfrac{1}{2}}$, $6p ^{2}P_{\sfrac{1}{2}} \to 6d ^{2}D_{\sfrac{3}{2}}$ and $6p ^{2}P_{\sfrac{1}{2}} \to 7s ^{2}S_{\sfrac{1}{2}}$ transitions in Tl.
}
\centering
\begin{tabular}{lccc}
\hline
\hline
 Transition & $6p ^{2}P_{\sfrac{3}{2}} \to 7s\ ^{2}S_{\sfrac{1}{2}}$(535~nm)  & ~~$6p ^{2}P_{\sfrac{1}{2}} \to 6d ^{2}D_{\sfrac{3}{2}}$(277~nm)  &   ~~$6p ^{2}P_{\sfrac{1}{2}} \to 7s ^{2}S_{\sfrac{1}{2}}$(378~nm)\\
\hline
\multicolumn{4}{c}{Contributions:}    \\ 
81$e$-CCSD(T) & 16.07 & 9.59 & 15.17 \\
~~21$e$-CCSDT$-$21$e$-CCSD(T) & $-$0.02 & $-$0.09 & $-$0.02 \\
~~21$e$-CCSDT(Q)$-$21$e$-CCSDT & $-$0.02 & $-$0.07 & $-$0.04 \\
~~Gaunt (FS-CCSD) & $-$0.11 & $-$0.08 & $-$0.10 \\
~~Breit (retardation part) & 0.01 & 0.01 & 0.01 \\
~~Charge distribution model & 0.22 & 0.13 & 0.21 \\
\multicolumn{4}{c}{Uncertainties:}    \\ 
Basis set
 & 0.11 & 0.39 & 0.05 \\
Correlation (core electrons)
 & 0.02 & 0.04 & 0.04 \\
Correlation (valence electrons)
 & 0.02 & 0.07 &  0.04\\
 Correlation and basis interference & 0.01 & 0.01 & 0.01 \\
 Breit & 0.01 & 0.01 & 0.01 \\
 Charge distribution model & 0.22 & 0.13 & 0.21 \\
 QED & 0.20 & 0.13 & 0.20 \\ 
\\ 
\textbf{Total} & 16.15(32) & 9.50(44) & 15.22(30) \\
\hline
\\
Other calculations & & &   \\ 
 SCDF$^a$~\cite{Hermann1993,Fricke18} & 17.58$^b$ & 11.90$^b$ & 16.69$^b$ \\
 SCDF$^a$~\cite{Menges1992,Fricke22} & 17.77$^c$ &  &  \\
 MBPT$^d$~\cite{Martensson1991} & 20.75$^c$ &  & 20.66$^c$ \\
 CC$^e$~\cite{Martensson1995} & 15.65 &  & 14.35  \\
 \hline
 \\
 Empirical evaluations & & & \\
 
 Ref.~\onlinecite{Fricke2004NuclearCR} & 16.88$^c$ &  &  \\
 Ref.~\onlinecite{Martensson1995} & 15.97(63) &  & 15.17(63) \\
\hline
\hline
\end{tabular}
\begin{flushleft}
    $^a$Single-configuration Dirac-Fock method. \\
    $^b$The values of the $F'$ factors from Table~5 of Ref.~\onlinecite{Hermann1993} are given here as they correspond to the $F$ factor in Eq.~(\ref{freq}) of the present paper. \\
    $^{c}$These electronic factors were obtained by multiplying the ``$F$ '' factor value from the corresponding references by the factor of  0.938~\cite{Fricke2004NuclearCR} due to a different definition of the $F$ factor in those references and in the present paper. \\
    $^d$Many-body perturbation theory, see Ref.~\onlinecite{Martensson1991} for details. \\
    $^e$A variant of the coupled cluster with single and double cluster amplitudes method, see Ref.~\onlinecite{Martensson1995} for details.\\
\end{flushleft}
\label{table:1}
\end{table*}

For correlation calculations, the finite-field technique was applied. The relativistic electronic structure calculations were performed using the locally modified versions of {\sc dirac}~\cite{DIRAC19,Saue:2020}, {\sc mrcc}~\cite{MRCC2020,Kallay:1,Kallay:2} and Exp-T~\cite{Oleynichenko_EXPT,EXPT_website} codes.

Note, that the nuclear structure effects such as the nuclear deformation and polarization~\cite{Yang2022_exotic,Zubova:2014,Wansbeek2012} can lead to additional terms in Eq.~(\ref{freq}). 
In the present paper, we do not consider such terms.

\section{Results and discussion}

Calculated values of the $F$, $k_{\rm NMS}$, $k_{\rm SMS}$, and $k_{\rm MS} = k_{\rm NMS} + k_{\rm SMS}$ factors and their uncertainties for three considered atomic transitions are given in Tables~\ref{table:1} and~\ref{k_all}. 
It can be seen from Table~\ref{table:1} that electronic correlation effects beyond the CCSD(T)  model are small for the $F$ factor. The Gaunt interelectron interaction contributes less than 1\% to the $F$ factor, whereas the influence of the retardation part of the zero-frequency Breit interaction is almost an order of magnitude smaller. The nuclear charge distribution correction is about 1\%. Calculation of mass shift constants, especially $k_{\rm SMS}$, is more challenging (see Table~\ref{k_all}). The connected quadruple cluster amplitudes' contribution calculated as the difference between the values obtained at the CCSDT(Q) and CCSDT levels [see line ``21$e$-CCSDT(Q)$-$21$e$-CCSDT''] is non-negligible for the $k_{\rm SMS}$ constant, whereas for $k_{\rm NMS}$, it is almost negligible for all transitions. The Gaunt-interaction correction, estimated for both $k_{\rm NMS}$ and $k_{\rm SMS}$ is non-negligible. Note, however, that Gaunt contributions to NMS and SMS factors for $6p ^{2}P_{\sfrac{3}{2}} \to 7s ^{2}S_{\sfrac{1}{2}}$ and $6p ^{2}P_{\sfrac{1}{2}} \to 7s ^{2}S_{\sfrac{1}{2}}$ transitions almost cancel each other and modestly contribute to the $k_{\rm MS}$ values for these transitions.

\begin{table*}
\caption{\label{k_all} Calculated values and theoretical uncertainties of $k_{\rm NMS}$ and $k_{\rm SMS}$$^a$ (in GHz$\cdot$u) and their sum $k_{\rm MS}$ for the $6p ^{2}P_{\sfrac{3}{2}} \to 7s ^{2}S_{\sfrac{1}{2}}$, $6p ^{2}P_{\sfrac{1}{2}} \to 6d ^{2}D_{\sfrac{3}{2}}$ and $6p ^{2}P_{\sfrac{1}{2}} \to 7s ^{2}S_{\sfrac{1}{2}}$ transitions in Tl. }
\begin{ruledtabular}
\begin{tabular}{l ccc ccc ccc}
 Transition & \multicolumn{3}{c}{$6p ^{2}P_{\sfrac{3}{2}} \to 7s ^{2}S_{\sfrac{1}{2}}$(535~nm)} & \multicolumn{3}{c}{$6p ^{2}P_{\sfrac{1}{2}} \to 6d ^{2}D_{\sfrac{3}{2}}$(277~nm)} & \multicolumn{3}{c}{$6p ^{2}P_{\sfrac{1}{2}} \to 7s ^{2}S_{\sfrac{1}{2}}$(378~nm)}\\
 \hline
 & $k_{\rm NMS}$ & $k_{\rm SMS}$ & $k_{\rm MS}$ & $k_{\rm NMS}$ & $k_{\rm SMS}$ & $k_{\rm MS}$ & $k_{\rm NMS}$ & $k_{\rm SMS}$ & $k_{\rm MS}$ \\
 \cline{2-4} \cline{5-7} \cline{8-10}
& \multicolumn{9}{c}{Contributions:} \\ 
 81$e$-CCSD(T)
 & $-$322 & 185 & $-$137 & $-$586 & $-$33 & $-$619 & $-$437 & 144 & $-$293 \\
 ~~21$e$-CCSDT$-$21$e$-CCSD(T)
 & $-$10 & 12 & 2 & $-$13 & 15 & 2 & $-$14 & 12 & $-$1 \\
 ~~21$e$-CCSDT(Q)$-$21$e$-CCSDT
 & 1 & 13 & 13 & 4 & 17 & 20 & 4 & 26 & 30\\
 Basis set correction
 & -- & $-$12 & $-$12 & -- & 54 & 54 & -- & $-$15 & $-$15 \\
 Gaunt (SCF)
 & 8 & $-$9 & $-$2 & $-$10 & $-$22 & $-$32 & 26 & $-$23 & 4\\
 & \multicolumn{9}{c}{Uncertainties:} \\
Basis set
 & 1 & 12 &  & 7 & 54 &  & 0 & 15 &  \\
Correlation (core electrons)
 & 5 & 26 &  & 4 & 30 &  & 5 & 22 &  \\
 Correlation (valence electrons)
 & 1 & 13 &  & 4 & 17 &  & 4 & 26 &  \\
 Correlation and basis interference & 3 & 12 &  & 1 & 15 &  & 3 & 15 &  \\
 Gaunt & 8 & 9 &  & 10 & 22 &  & 26 & 23 &  \\
\\ 
 \textbf{Total} & $-$323(10)  & 188(35) & $-$135(36) & $-$605(14) & 30(69) & $-$575(71) & $-$421(27) & 145(46) & $-$275(54) \\
\hline 
\\
Other calculations & & & & & & & & & \\
MBPT$^b$~\cite{Martensson1991} &  & $-$330$^c$ & &  & & & & 160$^c$ &  \\
HF$^d$~\cite{Wilson22} & $-$374 & 333 & &  &  & & & &  \\ 
\end{tabular}
\begin{flushleft}
$^a$In addition, we calculated the values of mass shift 
 for $^{203,205}$Tl ($\Delta\nu^{203,205}_{\rm MS}$) to illustrate the contribution of mass shift to IS. The obtained values are $-6.5$ MHz, $-27.5$ MHz and $-13.2$ MHz for $6p ^{2}P_{\sfrac{3}{2}} \to 7s ^{2}S_{\sfrac{1}{2}}$, $6p ^{2}P_{\sfrac{1}{2}} \to 6d ^{2}D_{\sfrac{3}{2}}$ and $6p ^{2}P_{\sfrac{1}{2}} \to 7s ^{2}S_{\sfrac{1}{2}}$ transitions, respectively. \\
 $^b$Many-body perturbation theory, see Ref.~\onlinecite{Martensson1991} for details. \\
$^c$ In Ref.~\onlinecite{Martensson1991} the authors state that obtained theoretical values of $k_{\rm SMS}$ are unreliable. They eventually prefer to use estimation $k_{\rm SMS} = (0 \pm 1.5) \times k_{\rm NMS}$. \\
$^d$``Hartree-Fock'' method~\cite{Wilson22}. \\
\end{flushleft}
\end{ruledtabular}
\end{table*}

The following sources of theoretical uncertainties of calculated isotope shift factors were considered.
(i)~The uncertainty due to the basis set incompleteness for $F$ and $k_{\rm NMS}$ factors was estimated as a difference between the values obtained at the CCSD(T) level using the LBas and MBas basis sets. For $k_{\rm SMS}$, this uncertainty was estimated at the CCSD level as a difference between the values obtained within the Lbas and MBasExt basis sets. This uncertainty contribution is given in the ``Basis set'' line of Tables~\ref{table:1} and \ref{k_all}.
(ii) The uncertainty due to the incomplete treatment of correlation effects of the sixty inner-core $1s...4f$ electrons of Tl was estimated as the difference between contributions of non-iterative triple cluster amplitudes for all 81 electrons and for 21 valence electrons. This difference corresponds to the contribution of the triple cluster amplitudes of the inner-core electrons to the considered factors. We suggest that the higher-order correlation effects for the inner-core electrons can contribute on the same order of magnitude. This uncertainty contribution is given in the ``Correlation (core electrons)'' line of Tables~\ref{table:1} and \ref{k_all}. 
(iii) The uncertainty due to the treatment of correlation effects beyond the CCSDT(Q) model for 21 valence and outer-core electrons was estimated as the difference between CCSDT(Q) and CCSDT results for 21-electron correlation calculations in the MBas basis set. This uncertainty contribution is given in the ``Correlation (valence electrons)'' line of Tables~\ref{table:1} and \ref{k_all}.
(iv) We also estimated the effect of the interference between the basis set size and the high-order correlation effects (i.e., correlation effects beyond the CCSD(T) model) for the valence and outer-core electrons. For this, we compared contributions of high-order correlation effects calculated within the MBas basis set and SBas basis set. This uncertainty contribution is given in the ``Correlation and basis interference'' line of Tables~\ref{table:1} and \ref{k_all}.
(v) The total value of the retardation part of the Breit interaction contribution given in Table~\ref{table:1} and the total value of the Gaunt interaction contribution given in Table~\ref{k_all} were considered as uncertainties (i.e., we assumed conservatively that these corrections have uncertainties of 100\%). 
(vi) The total charge distribution model correction from Table~\ref{table:1} was treated as the uncertainty of the FS.
(vii) Finally, we estimated one more source of the uncertainty for the field shift constant -- contribution of the quantum electrodynamics (QED) effects. According to our estimation, within the technique developed in Ref.~\onlinecite{Skripnikov:2021a}, the effect of the vacuum polarization (in the Uehling potential approximation) can contribute 1.3\% to the value of the FS constant. No rigorous quantum-electrodynamics calculation of the self-energy contribution to the FS constant is available for neutral thallium. However, according to Ref.~\onlinecite{Yerokhin:2011b}, there is a strong cancellation of the nuclear-size vacuum polarization and self-energy quantum electrodynamics corrections in hydrogen-like ions in the vicinity of Z=82.
The total contribution of QED effects to the FS constant of lithium-like Th was found to be about 0.5\%-0.7\%~\cite{Zubova:2014}.
We suggest that a similar effect should approximately hold for the present case. 
Thus, we estimate that the unaccounted QED effects for the neutral thallium case should not exceed the estimated vacuum-polarization contribution. We include it in the uncertainty. The total uncertainty for each constant was calculated as a square root of the sum of squares of uncertainties (i)-(vii).

As one can see from Table~\ref{k_all}, for all considered transitions, the absolute values of $k_{\rm SMS}$ are smaller than that for the $k_{\rm NMS}$. Nevertheless, the leading source of the uncertainty of $k_{\rm MS}$ is determined by the uncertainty of $k_{\rm SMS}$. The uncertainty of $k_{\rm SMS}$ for $6p ^{2}P_{\sfrac{1}{2}} \to 7s ^{2}S_{\sfrac{1}{2}}$ and $6p ^{2}P_{\sfrac{3}{2}} \to 7s ^{2}S_{\sfrac{1}{2}}$ transitions is dominated by the incomplete correlation effects treatment. For the $6p ^{2}P_{\sfrac{1}{2}} \to 6d ^{2}D_{\sfrac{3}{2}}$ transition, the uncertainties due to incomplete treatment of correlation effects and use of an incomplete basis set lead to the large relative uncertainty of $k_{\rm SMS}$ for this transition. However, the absolute value of $k_{\rm SMS}$ is found to be much smaller than the absolute value of $k_{\rm NMS}$ for this transition. Thus, the relative uncertainty of the total mass shift factor $k_{\rm MS}$ for the $6p ^{2}P_{\sfrac{1}{2}} \to 6d ^{2}D_{\sfrac{3}{2}}$ transition is similar to the uncertainties of $k_{\rm MS}$ for other two transitions.

It can be seen from Table~\ref{table:1}, that the obtained values of the $F$ factor for $6p ^{2}P_{\sfrac{3}{2}} \to 7s ^{2}S_{\sfrac{1}{2}}$ and $6p ^{2}P_{\sfrac{1}{2}} \to 7s ^{2}S_{\sfrac{1}{2}}$ transitions are in reasonable agreement with theoretical values and empirical evaluations from Ref.~\onlinecite{Martensson1995}. Note that NMS and SMS contributions to the total mass shift constant only partly cancel each other for the $6p ^{2}P_{3/2} \to 7s ^{2}S_{1/2}$ transition in contrast to the suggestion made in Ref.~\onlinecite{Menges1992} based on the simple ``Hartree-Fock'' estimates.

To check the calculation results and their uncertainties, we compared the theoretical ratio $k=F_{277~{\rm nm}}/F_{535~{\rm nm}}$ with its experimental value obtained by the King-plot analysis~\cite{Barzakh2013}: $k_{\rm theor}=0.588(29)$, $k_{\rm expt}=0.577(9)$. Good agreement testifies to the correctness of the theoretical procedure. Note that in the framework of the single-configuration Dirac-Fock method, $k_{\rm theor} = 0.677$~\cite{Fricke18}. King plot also gives a possibility to compare the experimental and theoretical values of the following combination of the MS and FS constants,
\begin{equation}
   s = k_{\rm MS}^{277~\mathrm{nm}} - \frac{F_{\mathrm{277~nm}}}{F_{\mathrm{535~nm}}} k_{\rm MS}^{535~\mathrm{nm}}. 
\end{equation}
The theoretical result $s_{\rm theor} = -490(70)$ ${\rm GHz} \cdot \rm u$ is also in good agreement with the King-plot value $s_{\rm expt} = -540(330)$ ${\rm GHz} \cdot \rm u$~\cite{Barzakh2013}.

Our results enable one to recalculate all $\delta \langle r^2 \rangle$ values previously extracted from the IS data for the atomic transitions studied in the present work. In the case of 277~nm transition, there are two options: either one can use a purely theoretical $F$~factor~[9.50(44)~GHz/fm$^2$] or take into account more accurate King-plot $F$ factors ratio, i.e., use $F_{ \rm 277~nm}=F_{\rm 535~nm} k_{\rm expt}=9.32(23)$ GHz/fm$^2$. We preferred the second option since it gives lower uncertainties for the $\delta \langle r^2 \rangle$ values. 
Results of this recalculation are presented in Table~\ref{radius_uni}.
The experimental uncertainties are shown in parentheses. Curly brackets denote the uncertainty due to the theoretical indeterminacy of $F$, $k_{\rm NMS}$ and $k_{\rm SMS}$. For comparison, we also provide the values of $\delta \langle r^2 \rangle$ obtained in Refs.~\onlinecite{Barzakh2013,Barzakh2017} based on the  previously adopted FS and MS constants.

\begin{table}
\caption{\label{radius_uni} Combined set of $\delta \langle r^2 \rangle$ values for the Tl isotopic chains, extracted from IS data for all three considered transitions in Ref.~\onlinecite{Barzakh2013,Barzakh2017} and in the present paper. The first uncertainty of the present $\delta \langle r^2 \rangle$ value is due to the experimental IS uncertainty and the second one given in the curly brackets is the theoretical one. If there are different data on same isotope, the one with the smallest total uncertainty is presented in this table. 
The nuclear structure-dependent effects were not considered.
}
\begin{ruledtabular}
\begin{tabular}{lccc}
A  & I & $\delta \langle r^2 \rangle$, fm$^2$~\cite{Barzakh2013,Barzakh2017} & $\delta \langle r^2 \rangle$, fm$^2$  \\
\hline
208g & 5 & 0.183(13)\{13\} & 0.1919(130)\{38\}$^{a}$ \\
207g & 1/2 & 0.1048(2)\{70\} & 0.1100(2)\{22\}$^{b}$ \\
205g & 1/2 & 0 & 0 \\
204g & 2 & $-$0.0635(71)\{40\} & $-$0.0667(74)\{13\}$^c$ \\
203g & 1/2 & $-$0.10321(2)\{700\} & $-$0.10840(3)\{220\}$^d$ \\
202g & 2 & $-$0.1834(71)\{130\} & $-$0.1926(74)\{38\}$^c$ \\
201g & 1/2 & $-$0.2077(9)\{150\} & $-$0.2182(9)\{43\}$^e$ \\
200g & 2 & $-$0.2979(71)\{210\} & $-$0.3129(74)\{62\}$^c$ \\
199g & 1/2 & $-$0.3116(71)\{220\} & $-$0.3275(74)\{65\}$^c$ \\
198g & 2 & $-$0.4035(71)\{290\} & $-$0.4239(74)\{84\}$^f$ \\
198m & 7 & $-$0.3804(71)\{270\} & $-$0.3998(74)\{80\}$^g$ \\
197g & 1/2 & $-$0.4119(71)\{290\} & $-$0.4330(74)\{86\}$^f$ \\
197m & 9/2 & $-$0.272(26)\{19\} & $-$0.2871(270)\{75\}$^h$ \\
196g & 2 & $-$0.4795(5)\{340\} & $-$0.5036(5)\{100\}$^i$ \\
196m & 7 & $-$0.4544(6)\{320\} & $-$0.4773(6)\{95\}$^i$ \\
195g & 1/2 & $-$0.4820(71)\{340\} & $-$0.5068(75)\{100\}$^j$ \\
195m & 9/2 & $-$0.324(11)\{23\} & $-$0.3419(120)\{90\}$^h$ \\
194g & 2 & $-$0.5551(39)\{50\} & $-$0.5831(5)\{120\}$^i$ \\
194m & 7 & $-$0.5481(5)\{380\} & $-$0.5759(5)\{110\}$^i$ \\
193g & 1/2 & $-$0.5716(11)\{400\} & $-$0.6007(12)\{120\}$^e$ \\
193m & 9/2 & $-$0.4111(10)\{290\} & $-$0.4329(11)\{87\}$^e$ \\
192g & 2 & $-$0.6296(4)\{440\} & $-$0.6616(4)\{130\}$^i$ \\
192m & 7 & $-$0.6358(6)\{450\} & $-$0.6681(6)\{130\}$^i$ \\
191g & 1/2 & $-$0.6544(7)\{460\} & $-$0.6878(7)\{140\}$^i$ \\
191m & 9/2 & $-$0.4899(6)\{340\} & $-$0.5158(6)\{100\}$^i$ \\
190g & 2 & $-$0.7063(4)\{490\} & $-$0.7424(4)\{150\}$^i$ \\
190m & 7 & $-$0.7223(5)\{510\} & $-$0.7591(5)\{150\}$^i$ \\
189m & 9/2 & $-$0.5543(41)\{390\} & $-$0.5837(43)\{120\}$^e$ \\
188m & 7 & $-$0.8134(5)\{570\} & $-$0.8549(5)\{170\}$^i$ \\
187m & 9/2 & $-$0.616(31)\{43\} & $-$0.650(32)\{17\}$^h$ \\
186m1 & 7 & $-$0.9324(15)\{650\} & $-$0.9799(15)\{200\}$^k$ \\
186m2 & 10 & $-$0.719(23)\{50\} & $-$0.758(24)\{20\}$^h$ \\
185g & 1/2 & $-$0.938(41)\{66\} & $-$0.987(43)\{25\}$^h$ \\
185m & 9/2 & $-$0.731(29)\{51\} & $-$0.770(30)\{20\}$^h$ \\
184m1 & 2 & $-$0.979(32)\{69\} & $-$1.031(32)\{27\}$^l$ \\
184m2 & 7 & $-$0.976(24)\{68\} & $-$1.027(26)\{27\}$^l$ \\
184m3 & 10 & $-$0.777(20)\{54\} & $-$0.820(23)\{21\}$^l$ \\
183g & 1/2 & $-$1.033(15)\{72\} & $-$1.086(17)\{28\}$^l$ \\
183m & 9/2 & $-$0.775(15)\{54\} & $-$0.818(17)\{22\}$^l$ \\
182m1 & 4 & $-$1.120(18)\{78\} & $-$1.179(19)\{30\}$^l$ \\
182m2 & 7 & $-$1.123(30)\{78\} & $-$1.182(33)\{30\}$^l$ \\
181 & 1/2 & $-$1.174(16)\{82\} & $-$1.236(17)\{32\}$^l$ \\
180 & 4 & $-$1.254(22)\{88\} & $-$1.319(24)\{34\}$^l$ \\
179 & 1/2 & $-$1.274(29)\{89\} & $-$1.340(31)\{35\}$^l$ \\
\end{tabular}
\end{ruledtabular}
\begin{flushleft}
   $^a$IS data from Ref.~\onlinecite{PhysRevLett.68.1675} \\
   $^b$IS data from Ref.~\onlinecite{PhysRevLett.55.1559} \\
   $^c$IS data from Ref.~\onlinecite{Hull:61} \\
   $^d$IS data from Ref.~\onlinecite{Hermann1993} \\
   $^e$IS data from Ref.~\onlinecite{PhysRevC.36.2560} \\
   $^f$IS data from Ref.~\onlinecite{Davis:66} \\
   $^g$IS data from Ref.~\onlinecite{GOORVITCH19671} \\
   $^h$IS data from Ref.~\onlinecite{Barzakh2013} \\
   $^i$IS data from Ref.~\onlinecite{Menges1992} \\
   $^j$IS data from Ref.~\onlinecite{PhysRev.188.1897} \\
   $^k$IS data from Ref.~\onlinecite{SCHUESSLER1995583} \\
   $^l$IS data from Ref.~\onlinecite{Barzakh2017} \\
\end{flushleft}
\end{table}

One can see that for the most exotic isotopes (A$<$186), the theoretical uncertainties become comparable with the experimental ones. Overall, the theoretical uncertainties are less than 2.6\%. This accuracy is at least not worse than the accuracy of the recent theoretical analysis of IS in radium~\cite{Wansbeek2012} and francium~\cite{Kalita2018} when the heavy atoms were taken for the comparison.
The difference between the literature and recalculated values is of the order of $~$5\%. Formally, this difference is covered by the theoretical uncertainties ascribed to the $\delta \langle r^2 \rangle$ values in Refs.~\onlinecite{Barzakh2013,Barzakh2017}. However, it should be reminded that the estimation of the theoretical uncertainty in Refs.~\onlinecite{Barzakh2013,Barzakh2017} (7\%) was based on rather general assumptions and could not be verified specifically for thallium without calculations similar to that accomplished in the present work. Moreover, the large difference between various calculations (up to 30\%, see Table~\ref{table:1}), pointed to the possibility of a much more pessimistic scenario.
In the procedure that was used in Refs.~\onlinecite{Barzakh2013,Barzakh2017} to extract the values of $\delta \langle r^2 \rangle$, the specific mass shift constant for the $6p ^{2}P_{\sfrac{3}{2}} \to 7s ^{2}S_{\sfrac{1}{2}}$ transition was neglected. However, as can be seen from Table~\ref{k_all}, $k_{\rm SMS}$ is not negligible for this transition and should be taken into account.

\section{Conclusions}
We performed the calculation of electronic factors of IS for Tl taking into account relativistic and high-order electronic correlation effects. The latter were considered within the coupled cluster theory with the inclusion of up to quadruple cluster amplitudes. As far as we know, these types of high-order correlation effects were not considered previously for isotope shifts in neutral heavy atoms. The developed procedure allowed us to update values of ms charge radii for plenty of thallium isotopes. Theoretical uncertainties of $\delta \langle r^2 \rangle$ were systematically estimated for the first time for the thallium isotopes. They were substantially reduced compared with the previously ascribed ones.
For the most exotic isotopes, theoretical uncertainties become comparable with the experimental ones. The achieved accuracy paves the way for a more reliable comparison of the charge radii trends in the lead region.

We did not consider the nuclear structure-dependent effects such as nuclear deformation and polarization. According to Ref.~\onlinecite{Moller:2016}, the static nuclear deformation of Tl isotopes is expected to be small. 
Taking into account previous investigations of the effects (see e.g.,~\onlinecite{Zubova:2014,Wansbeek2012}), we estimated that their contribution should be less than 2\% for Tl.
%Thus, the achieved accuracy in accounting for the electronic effects puts on the order of the day the rigorous treatment of the nuclear-structure effects contribution in the IS values for the heavy atoms.
Thus, the achieved accuracy in accounting for the electronic effects calls for the state-of-the-art treatment of the nuclear-structure effects contribution in the IS values for heavy atoms.

\section{Acknowledgments}
Electronic structure calculations in the paper were carried out using resources of the collective usage center ``Modeling and Predicting Properties of Materials'' at NRC ``Kurchatov Institute'' - PNPI.
The research has been supported by the Russian Science Foundation Grant No. 19-72-10019. 
Calculations of the electronic energies have been supported by the Foundation for the Advancement of Theoretical Physics and Mathematics ``BASIS'' grant according to the Research Project No. 21-1-2-47-1.

\section*{Author Declarations}
\subsection*{Conflict of interest}
The authors have no conflicts to disclose.

\section*{DATA AVAILABILITY}The data that support the findings of this study are available from the corresponding author upon reasonable request.


\begin{thebibliography}{67}
\expandafter\ifx\csname natexlab\endcsname\relax\def\natexlab#1{#1}\fi
\expandafter\ifx\csname bibnamefont\endcsname\relax
  \def\bibnamefont#1{#1}\fi
\expandafter\ifx\csname bibfnamefont\endcsname\relax
  \def\bibfnamefont#1{#1}\fi
\expandafter\ifx\csname citenamefont\endcsname\relax
  \def\citenamefont#1{#1}\fi
\expandafter\ifx\csname url\endcsname\relax
  \def\url#1{\texttt{#1}}\fi
\expandafter\ifx\csname urlprefix\endcsname\relax\def\urlprefix{URL }\fi
\providecommand{\bibinfo}[2]{#2}
\providecommand{\eprint}[2][]{\url{#2}}

\bibitem[{\citenamefont{Yang et~al.}(2022)\citenamefont{Yang, Wang, Wilkins,
  and Ruiz}}]{Yang2022_exotic}
\bibinfo{author}{\bibfnamefont{X.~F.}~\bibnamefont{Yang}},
  \bibinfo{author}{\bibfnamefont{S.~J.}~\bibnamefont{Wang}},
  \bibinfo{author}{\bibfnamefont{S.~G.}~\bibnamefont{Wilkins}}, \bibnamefont{and}
  \bibinfo{author}{\bibfnamefont{R.~F.~G.} \bibnamefont{Ruiz}},
  \bibinfo{journal}{Prog. Part. Nucl. Phys.} \textbf{\bibinfo{volume}{129}}, \bibinfo{pages}{104005}
  (\bibinfo{year}{2022}).

\bibitem[{\citenamefont{Garcia~Ruiz et~al.}(2016)\citenamefont{Garcia~Ruiz,
  Bissell, Blaum, Ekstr{\"o}m, Fr{\"o}mmgen, Hagen, Hammen, Hebeler, Holt,
  Jansen et~al.}}]{Garcia2016}
\bibinfo{author}{\bibfnamefont{R.~F.} \bibnamefont{Garcia~Ruiz}},
  \bibinfo{author}{\bibfnamefont{M.~L.}~\bibnamefont{Bissell}},
  \bibinfo{author}{\bibfnamefont{K.}~\bibnamefont{Blaum}},
  \bibinfo{author}{\bibfnamefont{A.}~\bibnamefont{Ekstr{\"o}m}},
  \bibinfo{author}{\bibfnamefont{N.}~\bibnamefont{Fr{\"o}mmgen}},
  \bibinfo{author}{\bibfnamefont{G.}~\bibnamefont{Hagen}},
  \bibinfo{author}{\bibfnamefont{M.}~\bibnamefont{Hammen}},
  \bibinfo{author}{\bibfnamefont{K.}~\bibnamefont{Hebeler}},
  \bibinfo{author}{\bibfnamefont{J.~D.}~\bibnamefont{Holt}},
  \bibinfo{author}{\bibfnamefont{G.~R.}~\bibnamefont{Jansen}},
  \bibnamefont{et~al.}, \bibinfo{journal}{Nat. Phys.}
  \textbf{\bibinfo{volume}{12}}, \bibinfo{pages}{594} (\bibinfo{year}{2016}).

\bibitem[{\citenamefont{De~Groote et~al.}(2020)\citenamefont{De~Groote,
  Billowes, Binnersley, Bissell, Cocolios, Day~Goodacre, Farooq-Smith, Fedorov,
  Flanagan, Franchoo et~al.}}]{deGroote2020}
\bibinfo{author}{\bibfnamefont{R.~P.}~\bibnamefont{De~Groote}},
  \bibinfo{author}{\bibfnamefont{J.}~\bibnamefont{Billowes}},
  \bibinfo{author}{\bibfnamefont{C.~L.} \bibnamefont{Binnersley}},
  \bibinfo{author}{\bibfnamefont{M.~L.} \bibnamefont{Bissell}},
  \bibinfo{author}{\bibfnamefont{T.~E.} \bibnamefont{Cocolios}},
  \bibinfo{author}{\bibfnamefont{T.}~\bibnamefont{Day~Goodacre}},
  \bibinfo{author}{\bibfnamefont{G.~J.} \bibnamefont{Farooq-Smith}},
  \bibinfo{author}{\bibfnamefont{D.~V.}~\bibnamefont{Fedorov}},
  \bibinfo{author}{\bibfnamefont{K.~T.} \bibnamefont{Flanagan}},
  \bibinfo{author}{\bibfnamefont{S.}~\bibnamefont{Franchoo}},
  \bibnamefont{et~al.}, \bibinfo{journal}{Nat. Phys.}
  \textbf{\bibinfo{volume}{16}}, \bibinfo{pages}{620} (\bibinfo{year}{2020}).

\bibitem[{\citenamefont{Koszor{\'u}s et~al.}(2021)\citenamefont{Koszor{\'u}s,
  Yang, Jiang, Novario, Bai, Billowes, Binnersley, Bissell, Cocolios, Cooper
  et~al.}}]{Koszorus2021}
\bibinfo{author}{\bibfnamefont{{\'A}.}~\bibnamefont{Koszor{\'u}s}},
  \bibinfo{author}{\bibfnamefont{X.~F.}~\bibnamefont{Yang}},
  \bibinfo{author}{\bibfnamefont{W.~G.}~\bibnamefont{Jiang}},
  \bibinfo{author}{\bibfnamefont{S.~J.}~\bibnamefont{Novario}},
  \bibinfo{author}{\bibfnamefont{S.~W.}~\bibnamefont{Bai}},
  \bibinfo{author}{\bibfnamefont{J.}~\bibnamefont{Billowes}},
  \bibinfo{author}{\bibfnamefont{C.~L.}~\bibnamefont{Binnersley}},
  \bibinfo{author}{\bibfnamefont{M.~L.}~\bibnamefont{Bissell}},
  \bibinfo{author}{\bibfnamefont{T.~E.} \bibnamefont{Cocolios}},
  \bibinfo{author}{\bibfnamefont{B.~S.}~\bibnamefont{Cooper}},
  \bibnamefont{et~al.}, \bibinfo{journal}{Nat. Phys.}
  \textbf{\bibinfo{volume}{17}}, \bibinfo{pages}{439} (\bibinfo{year}{2021}).

\bibitem[{\citenamefont{Day~Goodacre et~al.}(2021)\citenamefont{Day~Goodacre,
  Afanasjev, Barzakh, Marsh, Sels, Ring, Nakada, Andreyev, Van~Duppen,
  Althubiti et~al.}}]{Goodacre2021}
\bibinfo{author}{\bibfnamefont{T.}~\bibnamefont{Day~Goodacre}},
  \bibinfo{author}{\bibfnamefont{A.~V.} \bibnamefont{Afanasjev}},
  \bibinfo{author}{\bibfnamefont{A.~E.} \bibnamefont{Barzakh}},
  \bibinfo{author}{\bibfnamefont{B.~A.} \bibnamefont{Marsh}},
  \bibinfo{author}{\bibfnamefont{S.}~\bibnamefont{Sels}},
  \bibinfo{author}{\bibfnamefont{P.}~\bibnamefont{Ring}},
  \bibinfo{author}{\bibfnamefont{H.}~\bibnamefont{Nakada}},
  \bibinfo{author}{\bibfnamefont{A.~N.} \bibnamefont{Andreyev}},
  \bibinfo{author}{\bibfnamefont{P.}~\bibnamefont{Van~Duppen}},
  \bibinfo{author}{\bibfnamefont{N.~A.} \bibnamefont{Althubiti}},
  \bibnamefont{et~al.}, \bibinfo{journal}{Phys. Rev. Lett.}
  \textbf{\bibinfo{volume}{126}}, \bibinfo{pages}{032502}
  (\bibinfo{year}{2021}).

\bibitem[{\citenamefont{Pineda et~al.}(2021)\citenamefont{Pineda, K\"onig,
  Rossi, Brown, Incorvati, Lantis, Minamisono, N\"ortersh\"auser, Piekarewicz,
  Powel et~al.}}]{Pineda2021}
\bibinfo{author}{\bibfnamefont{S.~V.} \bibnamefont{Pineda}},
  \bibinfo{author}{\bibfnamefont{K.}~\bibnamefont{K\"onig}},
  \bibinfo{author}{\bibfnamefont{D.~M.} \bibnamefont{Rossi}},
  \bibinfo{author}{\bibfnamefont{B.~A.} \bibnamefont{Brown}},
  \bibinfo{author}{\bibfnamefont{A.}~\bibnamefont{Incorvati}},
  \bibinfo{author}{\bibfnamefont{J.}~\bibnamefont{Lantis}},
  \bibinfo{author}{\bibfnamefont{K.}~\bibnamefont{Minamisono}},
  \bibinfo{author}{\bibfnamefont{W.}~\bibnamefont{N\"ortersh\"auser}},
  \bibinfo{author}{\bibfnamefont{J.}~\bibnamefont{Piekarewicz}},
  \bibinfo{author}{\bibfnamefont{R.}~\bibnamefont{Powel}}, and
  \bibinfo{author}{\bibfnamefont{F.}~\bibnamefont{Sommer}},
  \bibinfo{journal}{Phys. Rev. Lett.}
  \textbf{\bibinfo{volume}{127}}, \bibinfo{pages}{182503}
  (\bibinfo{year}{2021}).

\bibitem[{\citenamefont{Han et~al.}(2022)\citenamefont{Han, Pan, Zhang, Yang,
  Zhang, Berengut, Goriely, Wang, Yu, Meng et~al.}}]{Han2022}
\bibinfo{author}{\bibfnamefont{J.~Z.} \bibnamefont{Han}},
  \bibinfo{author}{\bibfnamefont{C.}~\bibnamefont{Pan}},
  \bibinfo{author}{\bibfnamefont{K.~Y.} \bibnamefont{Zhang}},
  \bibinfo{author}{\bibfnamefont{X.~F.} \bibnamefont{Yang}},
  \bibinfo{author}{\bibfnamefont{S.~Q.} \bibnamefont{Zhang}},
  \bibinfo{author}{\bibfnamefont{J.~C.} \bibnamefont{Berengut}},
  \bibinfo{author}{\bibfnamefont{S.}~\bibnamefont{Goriely}},
  \bibinfo{author}{\bibfnamefont{H.}~\bibnamefont{Wang}},
  \bibinfo{author}{\bibfnamefont{Y.~M.} \bibnamefont{Yu}},
  \bibinfo{author}{\bibfnamefont{J.}~\bibnamefont{Meng}}, \bibnamefont{et~al.},
  \bibinfo{journal}{Phys. Rev. Res.} \textbf{\bibinfo{volume}{4}},
  \bibinfo{pages}{033049} (\bibinfo{year}{2022}).

\bibitem[{\citenamefont{Barzakh et~al.}(2021)\citenamefont{Barzakh, Andreyev,
  Raison, Cubiss, Van~Duppen, P\'eru, Hilaire, Goriely, Andel, Antalic
  et~al.}}]{Barzakh2021}
\bibinfo{author}{\bibfnamefont{A.}~\bibnamefont{Barzakh}},
  \bibinfo{author}{\bibfnamefont{A.~N.} \bibnamefont{Andreyev}},
  \bibinfo{author}{\bibfnamefont{C.}~\bibnamefont{Raison}},
  \bibinfo{author}{\bibfnamefont{J.~G.} \bibnamefont{Cubiss}},
  \bibinfo{author}{\bibfnamefont{P.}~\bibnamefont{Van~Duppen}},
  \bibinfo{author}{\bibfnamefont{S.}~\bibnamefont{P\'eru}},
  \bibinfo{author}{\bibfnamefont{S.}~\bibnamefont{Hilaire}},
  \bibinfo{author}{\bibfnamefont{S.}~\bibnamefont{Goriely}},
  \bibinfo{author}{\bibfnamefont{B.}~\bibnamefont{Andel}},
  \bibinfo{author}{\bibfnamefont{S.}~\bibnamefont{Antalic}},
  \bibnamefont{et~al.}, \bibinfo{journal}{Phys. Rev. Lett.}
  \textbf{\bibinfo{volume}{127}}, \bibinfo{pages}{192501}
  (\bibinfo{year}{2021}).

\bibitem[{\citenamefont{Fricke and Heilig}(2004)}]{Fricke2004NuclearCR}
\bibinfo{author}{\bibfnamefont{G.}~\bibnamefont{Fricke}} \bibnamefont{and}
  \bibinfo{author}{\bibfnamefont{K.}~\bibnamefont{Heilig}},
  \emph{\bibinfo{title}{Nuclear Charge Radii}} (\bibinfo{publisher}{Springer,
  Berlin}, \bibinfo{year}{2004}).

\bibitem[{\citenamefont{Otten}(1989)}]{Otten1989}
\bibinfo{author}{\bibfnamefont{E.~W.} \bibnamefont{Otten}},
  \emph{\bibinfo{title}{Nuclear Radii and Moments of Unstable Isotopes}}
  (\bibinfo{publisher}{Springer US}, \bibinfo{address}{Boston, MA},
  \bibinfo{year}{1989}), pp. \bibinfo{pages}{517--638}, ISBN
  \bibinfo{isbn}{978-1-4613-0713-6}.

\bibitem[{\citenamefont{Papoulia et~al.}(2016)\citenamefont{Papoulia, Carlsson,
  and Ekman}}]{Papoulia2016}
\bibinfo{author}{\bibfnamefont{A.}~\bibnamefont{Papoulia}},
  \bibinfo{author}{\bibfnamefont{B.~G.} \bibnamefont{Carlsson}},
  \bibnamefont{and} \bibinfo{author}{\bibfnamefont{J.}~\bibnamefont{Ekman}},
  \bibinfo{journal}{Phys. Rev. A} \textbf{\bibinfo{volume}{94}},
  \bibinfo{pages}{042502} (\bibinfo{year}{2016}).

\bibitem[{\citenamefont{Kalita et~al.}(2018)\citenamefont{Kalita, Behr,
  Gorelov, Pearson, DeHart, Gwinner, Kossin, Orozco, Aubin, Gomez
  et~al.}}]{Kalita2018}
\bibinfo{author}{\bibfnamefont{M.~R.} \bibnamefont{Kalita}},
  \bibinfo{author}{\bibfnamefont{J.~A.} \bibnamefont{Behr}},
  \bibinfo{author}{\bibfnamefont{A.}~\bibnamefont{Gorelov}},
  \bibinfo{author}{\bibfnamefont{M.~R.} \bibnamefont{Pearson}},
  \bibinfo{author}{\bibfnamefont{A.~C.} \bibnamefont{DeHart}},
  \bibinfo{author}{\bibfnamefont{G.}~\bibnamefont{Gwinner}},
  \bibinfo{author}{\bibfnamefont{M.~J.} \bibnamefont{Kossin}},
  \bibinfo{author}{\bibfnamefont{L.~A.} \bibnamefont{Orozco}},
  \bibinfo{author}{\bibfnamefont{S.}~\bibnamefont{Aubin}},
  \bibinfo{author}{\bibfnamefont{E.}~\bibnamefont{Gomez}},
  \bibnamefont{et~al.}, \bibinfo{journal}{Phys. Rev. A}
  \textbf{\bibinfo{volume}{97}}, \bibinfo{pages}{042507}
  (\bibinfo{year}{2018}).

\bibitem[{\citenamefont{Sahoo et~al.}(2020)\citenamefont{Sahoo, Vernon, Ruiz,
  Binnersley, Billowes, Bissell, Cocolios, Farooq-Smith, Flanagan, Gins
  et~al.}}]{Sahoo_2020}
\bibinfo{author}{\bibfnamefont{B.~K.} \bibnamefont{Sahoo}},
  \bibinfo{author}{\bibfnamefont{A.~R.} \bibnamefont{Vernon}},
  \bibinfo{author}{\bibfnamefont{R.~F.} \bibnamefont{Garcia~Ruiz}},
  \bibinfo{author}{\bibfnamefont{C.~L.} \bibnamefont{Binnersley}},
  \bibinfo{author}{\bibfnamefont{J.}~\bibnamefont{Billowes}},
  \bibinfo{author}{\bibfnamefont{M.~L.} \bibnamefont{Bissell}},
  \bibinfo{author}{\bibfnamefont{T.~E.} \bibnamefont{Cocolios}},
  \bibinfo{author}{\bibfnamefont{G.~J.} \bibnamefont{Farooq-Smith}},
  \bibinfo{author}{\bibfnamefont{K.~T.} \bibnamefont{Flanagan}},
  \bibinfo{author}{\bibfnamefont{W.}~\bibnamefont{Gins}}, \bibnamefont{et~al.},
  \bibinfo{journal}{New J. Phys.} \textbf{\bibinfo{volume}{22}},
  \bibinfo{pages}{012001} (\bibinfo{year}{2020}).

\bibitem[{\citenamefont{Gustafsson et~al.}(2020)\citenamefont{Gustafsson,
  Ricketts, Reitsma, Garcia~Ruiz, Bai, Berengut, Billowes, Binnersley,
  Borschevsky, Cocolios et~al.}}]{Gustafsson2020}
\bibinfo{author}{\bibfnamefont{F.~P.} \bibnamefont{Gustafsson}},
  \bibinfo{author}{\bibfnamefont{C.~M.} \bibnamefont{Ricketts}},
  \bibinfo{author}{\bibfnamefont{M.~L.} \bibnamefont{Reitsma}},
  \bibinfo{author}{\bibfnamefont{R.~F.} \bibnamefont{Garcia~Ruiz}},
  \bibinfo{author}{\bibfnamefont{S.~W.} \bibnamefont{Bai}},
  \bibinfo{author}{\bibfnamefont{J.~C.} \bibnamefont{Berengut}},
  \bibinfo{author}{\bibfnamefont{J.}~\bibnamefont{Billowes}},
  \bibinfo{author}{\bibfnamefont{C.~L.} \bibnamefont{Binnersley}},
  \bibinfo{author}{\bibfnamefont{A.}~\bibnamefont{Borschevsky}},
  \bibinfo{author}{\bibfnamefont{T.~E.} \bibnamefont{Cocolios}},
  \bibnamefont{et~al.}, \bibinfo{journal}{Phys. Rev. A}
  \textbf{\bibinfo{volume}{102}}, \bibinfo{pages}{052812}
  (\bibinfo{year}{2020}).

\bibitem[{\citenamefont{Neugart et~al.}(1985)\citenamefont{Neugart, Stroke,
  Ahmad, Duong, Ravn, and Wendt}}]{PhysRevLett.55.1559}
\bibinfo{author}{\bibfnamefont{R.}~\bibnamefont{Neugart}},
  \bibinfo{author}{\bibfnamefont{H.~H.} \bibnamefont{Stroke}},
  \bibinfo{author}{\bibfnamefont{S.~A.} \bibnamefont{Ahmad}},
  \bibinfo{author}{\bibfnamefont{H.~T.} \bibnamefont{Duong}},
  \bibinfo{author}{\bibfnamefont{H.~L.} \bibnamefont{Ravn}}, \bibnamefont{and}
  \bibinfo{author}{\bibfnamefont{K.}~\bibnamefont{Wendt}},
  \bibinfo{journal}{Phys. Rev. Lett.} \textbf{\bibinfo{volume}{55}},
  \bibinfo{pages}{1559} (\bibinfo{year}{1985}).

\bibitem[{\citenamefont{Hull and Stroke}(1961)}]{Hull:61}
\bibinfo{author}{\bibfnamefont{R.~J.} \bibnamefont{Hull}} \bibnamefont{and}
  \bibinfo{author}{\bibfnamefont{H.~H.} \bibnamefont{Stroke}},
  \bibinfo{journal}{J. Opt. Soc. Am.} \textbf{\bibinfo{volume}{51}},
  \bibinfo{pages}{1203} (\bibinfo{year}{1961}).

\bibitem[{\citenamefont{Odintsov}(1960)}]{Odintsov1960}
\bibinfo{author}{\bibfnamefont{A.}~\bibnamefont{Odintsov}},
  \bibinfo{journal}{Opt. Spectrosc.} \textbf{\bibinfo{volume}{9}},
  \bibinfo{pages}{75} (\bibinfo{year}{1960}).

\bibitem[{\citenamefont{G.Hermann et~al.}(1993)\citenamefont{G.Hermann,
  Lasnitschka, and Spengler}}]{Hermann1993}
\bibinfo{author}{\bibnamefont{G.Hermann}},
  \bibinfo{author}{\bibfnamefont{G.}~\bibnamefont{Lasnitschka}},
  \bibnamefont{and} \bibinfo{author}{\bibfnamefont{D.}~\bibnamefont{Spengler}},
  \bibinfo{journal}{Z. Phys. D} \textbf{\bibinfo{volume}{28}},
  \bibinfo{pages}{127} (\bibinfo{year}{1993}).

\bibitem[{\citenamefont{Bounds et~al.}(1987)\citenamefont{Bounds, Bingham,
  Carter, Leander, Mlekodaj, Spejewski, and Fairbank}}]{PhysRevC.36.2560}
\bibinfo{author}{\bibfnamefont{J.~A.} \bibnamefont{Bounds}},
  \bibinfo{author}{\bibfnamefont{C.~R.} \bibnamefont{Bingham}},
  \bibinfo{author}{\bibfnamefont{H.~K.} \bibnamefont{Carter}},
  \bibinfo{author}{\bibfnamefont{G.~A.} \bibnamefont{Leander}},
  \bibinfo{author}{\bibfnamefont{R.~L.} \bibnamefont{Mlekodaj}},
  \bibinfo{author}{\bibfnamefont{E.~H.} \bibnamefont{Spejewski}},
  \bibnamefont{and} \bibinfo{author}{\bibfnamefont{W.~M.}
  \bibnamefont{Fairbank}}, \bibinfo{journal}{Phys. Rev. C}
  \textbf{\bibinfo{volume}{36}}, \bibinfo{pages}{2560} (\bibinfo{year}{1987}).

\bibitem[{\citenamefont{Goorvitch et~al.}(1967)\citenamefont{Goorvitch,
  Kleiman, and Davis}}]{GOORVITCH19671}
\bibinfo{author}{\bibfnamefont{D.}~\bibnamefont{Goorvitch}},
  \bibinfo{author}{\bibfnamefont{H.}~\bibnamefont{Kleiman}}, \bibnamefont{and}
  \bibinfo{author}{\bibfnamefont{S.~P.} \bibnamefont{Davis}},
  \bibinfo{journal}{Nucl. Phys.} \textbf{\bibinfo{volume}{99}},
  \bibinfo{pages}{1} (\bibinfo{year}{1967}).

\bibitem[{\citenamefont{Davis et~al.}(1966)\citenamefont{Davis, Kleiman,
  Goorvitch, and Aung}}]{Davis:66}
\bibinfo{author}{\bibfnamefont{S.~P.} \bibnamefont{Davis}},
  \bibinfo{author}{\bibfnamefont{H.}~\bibnamefont{Kleiman}},
  \bibinfo{author}{\bibfnamefont{D.}~\bibnamefont{Goorvitch}},
  \bibnamefont{and} \bibinfo{author}{\bibfnamefont{T.}~\bibnamefont{Aung}},
  \bibinfo{journal}{J. Opt. Soc. Am.} \textbf{\bibinfo{volume}{56}},
  \bibinfo{pages}{1604} (\bibinfo{year}{1966}).

\bibitem[{\citenamefont{Menges et~al.}(1992)\citenamefont{Menges, Dinger, Boos,
  Huber, Schröder, Dutta, Kirchner, Klepper, Kühl, Marx et~al.}}]{Menges1992}
\bibinfo{author}{\bibfnamefont{R.}~\bibnamefont{Menges}},
  \bibinfo{author}{\bibfnamefont{U.}~\bibnamefont{Dinger}},
  \bibinfo{author}{\bibfnamefont{N.}~\bibnamefont{Boos}},
  \bibinfo{author}{\bibfnamefont{G.}~\bibnamefont{Huber}},
  \bibinfo{author}{\bibfnamefont{S.}~\bibnamefont{Schröder}},
  \bibinfo{author}{\bibfnamefont{S.}~\bibnamefont{Dutta}},
  \bibinfo{author}{\bibfnamefont{R.}~\bibnamefont{Kirchner}},
  \bibinfo{author}{\bibfnamefont{O.}~\bibnamefont{Klepper}},
  \bibinfo{author}{\bibfnamefont{T.}~\bibnamefont{Kühl}},
  \bibinfo{author}{\bibfnamefont{D.}~\bibnamefont{Marx}}, and
  \bibinfo{author}{\bibfnamefont{G.~D.}~\bibnamefont{Sprouse}},
  \bibinfo{journal}{Z. Phys. A} \textbf{\bibinfo{volume}{341}},
  \bibinfo{pages}{475} (\bibinfo{year}{1992}).

\bibitem[{\citenamefont{Goorvitch et~al.}(1969)\citenamefont{Goorvitch, Davis,
  and Kleiman}}]{PhysRev.188.1897}
\bibinfo{author}{\bibfnamefont{D.}~\bibnamefont{Goorvitch}},
  \bibinfo{author}{\bibfnamefont{S.~P.} \bibnamefont{Davis}}, \bibnamefont{and}
  \bibinfo{author}{\bibfnamefont{H.}~\bibnamefont{Kleiman}},
  \bibinfo{journal}{Phys. Rev.} \textbf{\bibinfo{volume}{188}},
  \bibinfo{pages}{1897} (\bibinfo{year}{1969}).

\bibitem[{\citenamefont{Buchinger et~al.}(1992)\citenamefont{Buchinger,
  Schuessler, Benck, Iimura, Li, Bingham, and Carter}}]{Buchinger1992}
\bibinfo{author}{\bibfnamefont{F.}~\bibnamefont{Buchinger}},
  \bibinfo{author}{\bibfnamefont{H.~A.} \bibnamefont{Schuessler}},
  \bibinfo{author}{\bibfnamefont{E.~C.} \bibnamefont{Benck}},
  \bibinfo{author}{\bibfnamefont{H.}~\bibnamefont{Iimura}},
  \bibinfo{author}{\bibfnamefont{Y.~F.} \bibnamefont{Li}},
  \bibinfo{author}{\bibfnamefont{C.}~\bibnamefont{Bingham}}, \bibnamefont{and}
  \bibinfo{author}{\bibfnamefont{H.~K.} \bibnamefont{Carter}},
  \bibinfo{journal}{Hyperfine Interact.} \textbf{\bibinfo{volume}{75}},
  \bibinfo{pages}{367–371} (\bibinfo{year}{1992}).

\bibitem[{\citenamefont{Schuessler et~al.}(1995)\citenamefont{Schuessler,
  Benck, Buchinger, and Carter}}]{SCHUESSLER1995583}
\bibinfo{author}{\bibfnamefont{H.~A.}~\bibnamefont{Schuessler}},
  \bibinfo{author}{\bibfnamefont{E.~C.}~\bibnamefont{Benck}},
  \bibinfo{author}{\bibfnamefont{F.}~\bibnamefont{Buchinger}},
  \bibnamefont{and} \bibinfo{author}{\bibfnamefont{H.~K.}~\bibnamefont{Carter}},
  \bibinfo{journal}{Nucl. Instrum. Methods Phys. Res. A}
  \textbf{\bibinfo{volume}{352}}, \bibinfo{pages}{583} (\bibinfo{year}{1995}).

\bibitem[{\citenamefont{Barzakh et~al.}(2013)\citenamefont{Barzakh, Batist,
  Fedorov, Ivanov, Mezilev, Molkanov, Moroz, Orlov, Panteleev, and
  Volkov}}]{Barzakh2013}
\bibinfo{author}{\bibfnamefont{A.~E.} \bibnamefont{Barzakh}},
  \bibinfo{author}{\bibfnamefont{L.~K.} \bibnamefont{Batist}},
  \bibinfo{author}{\bibfnamefont{D.~V.} \bibnamefont{Fedorov}},
  \bibinfo{author}{\bibfnamefont{V.~S.} \bibnamefont{Ivanov}},
  \bibinfo{author}{\bibfnamefont{K.~A.} \bibnamefont{Mezilev}},
  \bibinfo{author}{\bibfnamefont{P.~L.} \bibnamefont{Molkanov}},
  \bibinfo{author}{\bibfnamefont{F.~V.} \bibnamefont{Moroz}},
  \bibinfo{author}{\bibfnamefont{S.~Y.} \bibnamefont{Orlov}},
  \bibinfo{author}{\bibfnamefont{V.~N.} \bibnamefont{Panteleev}},
  \bibnamefont{and} \bibinfo{author}{\bibfnamefont{Y.~M.}
  \bibnamefont{Volkov}}, \bibinfo{journal}{Phys. Rev. C}
  \textbf{\bibinfo{volume}{88}}, \bibinfo{pages}{024315}
  (\bibinfo{year}{2013}).

\bibitem[{\citenamefont{Barzakh et~al.}(2017)\citenamefont{Barzakh, Andreyev,
  Cocolios, de~Groote, Fedorov, Fedosseev, Ferrer, Fink, Ghys, Huyse
  et~al.}}]{Barzakh2017}
\bibinfo{author}{\bibfnamefont{A.~E.} \bibnamefont{Barzakh}},
  \bibinfo{author}{\bibfnamefont{A.~N.} \bibnamefont{Andreyev}},
  \bibinfo{author}{\bibfnamefont{T.~E.} \bibnamefont{Cocolios}},
  \bibinfo{author}{\bibfnamefont{R.~P.} \bibnamefont{de~Groote}},
  \bibinfo{author}{\bibfnamefont{D.~V.} \bibnamefont{Fedorov}},
  \bibinfo{author}{\bibfnamefont{V.~N.} \bibnamefont{Fedosseev}},
  \bibinfo{author}{\bibfnamefont{R.}~\bibnamefont{Ferrer}},
  \bibinfo{author}{\bibfnamefont{D.~A.} \bibnamefont{Fink}},
  \bibinfo{author}{\bibfnamefont{L.}~\bibnamefont{Ghys}},
  \bibinfo{author}{\bibfnamefont{M.}~\bibnamefont{Huyse}},
  \bibnamefont{et~al.}, \bibinfo{journal}{Phys. Rev. C}
  \textbf{\bibinfo{volume}{95}}, \bibinfo{pages}{014324}
  (\bibinfo{year}{2017}).

\bibitem[{\citenamefont{Lauth et~al.}(1992)\citenamefont{Lauth, Backe,
  Dahlinger, Klaft, Schwamb, Schwickert, Trautmann, and
  Othmer}}]{PhysRevLett.68.1675}
\bibinfo{author}{\bibfnamefont{W.}~\bibnamefont{Lauth}},
  \bibinfo{author}{\bibfnamefont{H.}~\bibnamefont{Backe}},
  \bibinfo{author}{\bibfnamefont{M.}~\bibnamefont{Dahlinger}},
  \bibinfo{author}{\bibfnamefont{I.}~\bibnamefont{Klaft}},
  \bibinfo{author}{\bibfnamefont{P.}~\bibnamefont{Schwamb}},
  \bibinfo{author}{\bibfnamefont{G.}~\bibnamefont{Schwickert}},
  \bibinfo{author}{\bibfnamefont{N.}~\bibnamefont{Trautmann}},
  \bibnamefont{and} \bibinfo{author}{\bibfnamefont{U.}~\bibnamefont{Othmer}},
  \bibinfo{journal}{Phys. Rev. Lett.} \textbf{\bibinfo{volume}{68}},
  \bibinfo{pages}{1675} (\bibinfo{year}{1992}).

\bibitem[{\citenamefont{Schuler et~al.}(1962)\citenamefont{Schuler, \c{C}iftan,
  Bradley, and Stroke}}]{Schuler:62}
\bibinfo{author}{\bibfnamefont{C.~J.} \bibnamefont{Schuler}},
  \bibinfo{author}{\bibfnamefont{M.}~\bibnamefont{\c{C}iftan}},
  \bibinfo{author}{\bibfnamefont{L.~C.} \bibnamefont{Bradley}},
  \bibnamefont{and} \bibinfo{author}{\bibfnamefont{H.~H.}
  \bibnamefont{Stroke}}, \bibinfo{journal}{J. Opt. Soc. Am.}
  \textbf{\bibinfo{volume}{52}}, \bibinfo{pages}{501} (\bibinfo{year}{1962}).

\bibitem[{\citenamefont{Richardson et~al.}(2000)\citenamefont{Richardson,
  Lyman, and Majumder}}]{PhysRevA.62.012510}
\bibinfo{author}{\bibfnamefont{D.~S.} \bibnamefont{Richardson}},
  \bibinfo{author}{\bibfnamefont{R.~N.} \bibnamefont{Lyman}}, \bibnamefont{and}
  \bibinfo{author}{\bibfnamefont{P.~K.} \bibnamefont{Majumder}},
  \bibinfo{journal}{Phys. Rev. A} \textbf{\bibinfo{volume}{62}},
  \bibinfo{pages}{012510} (\bibinfo{year}{2000}).

\bibitem[{\citenamefont{Marsh et~al.}(2018)\citenamefont{Marsh, Day~Goodacre,
  Sels, Tsunoda, Andel, Andreyev, Althubiti, Atanasov, Barzakh, Billowes
  et~al.}}]{Marsh2018}
\bibinfo{author}{\bibfnamefont{B.~A.}~\bibnamefont{Marsh}},
  \bibinfo{author}{\bibfnamefont{T.}~\bibnamefont{Day~Goodacre}},
  \bibinfo{author}{\bibfnamefont{S.}~\bibnamefont{Sels}},
  \bibinfo{author}{\bibfnamefont{Y.}~\bibnamefont{Tsunoda}},
  \bibinfo{author}{\bibfnamefont{B.}~\bibnamefont{Andel}},
  \bibinfo{author}{\bibfnamefont{A.~N.}~\bibnamefont{Andreyev}},
  \bibinfo{author}{\bibfnamefont{N.~A.}~\bibnamefont{Althubiti}},
  \bibinfo{author}{\bibfnamefont{D.}~\bibnamefont{Atanasov}},
  \bibinfo{author}{\bibfnamefont{A.~E.}~\bibnamefont{Barzakh}},
  \bibinfo{author}{\bibfnamefont{J.}~\bibnamefont{Billowes}},
  \bibnamefont{et~al.}, \bibinfo{journal}{Nat. Phys.}
  \textbf{\bibinfo{volume}{14}}, \bibinfo{pages}{1163} (\bibinfo{year}{2018}).

\bibitem[{\citenamefont{Angeli and Marinova}(2013)}]{ANGELI201369}
\bibinfo{author}{\bibfnamefont{I.}~\bibnamefont{Angeli}} \bibnamefont{and}
  \bibinfo{author}{\bibfnamefont{K.~P.}~\bibnamefont{Marinova}},
  \bibinfo{journal}{At. Data Nucl. Data Tables} \textbf{\bibinfo{volume}{99}},
  \bibinfo{pages}{69 } (\bibinfo{year}{2013}).

\bibitem[{\citenamefont{Shabaev}(1985)}]{shabaev1985mass}
\bibinfo{author}{\bibfnamefont{V.~M.} \bibnamefont{Shabaev}},
  \bibinfo{journal}{Theor. Math. Phys.} \textbf{\bibinfo{volume}{63}},
  \bibinfo{pages}{588} (\bibinfo{year}{1985}).

\bibitem[{\citenamefont{Palmer}(1987)}]{palmer1987reformulation}
\bibinfo{author}{\bibfnamefont{C.~W.~P.} \bibnamefont{Palmer}},
  \bibinfo{journal}{J. Phys. B: Atom. Mol. Phys.}
  \textbf{\bibinfo{volume}{20}}, \bibinfo{pages}{5987} (\bibinfo{year}{1987}).

\bibitem[{\citenamefont{Shabaev}(1988)}]{shabaev1988nucl}
\bibinfo{author}{\bibfnamefont{V.~M.} \bibnamefont{Shabaev}},
  \bibinfo{journal}{Sov. J. Nucl. Phys.} \textbf{\bibinfo{volume}{47}},
  \bibinfo{pages}{69} (\bibinfo{year}{1988}).

\bibitem[{\citenamefont{Shabaev and Artemyev}(1994)}]{shabaev1994relativistic}
\bibinfo{author}{\bibfnamefont{V.~M.}~\bibnamefont{Shabaev}} \bibnamefont{and}
  \bibinfo{author}{\bibfnamefont{A.~N.}~\bibnamefont{Artemyev}},
  \bibinfo{journal}{J. Phys. B: Atom. Mol. Phys.}
  \textbf{\bibinfo{volume}{27}}, \bibinfo{pages}{1307} (\bibinfo{year}{1994}).

\bibitem[{DIR()}]{DIRAC19}
\bibinfo{note}{DIRAC, a relativistic ab initio electronic structure program,
  Release DIRAC19 (2019), written by A. S. P. Gomes, T. Saue, L. Visscher, H.
  J. Aa. Jensen, and R. Bast, with contributions from I. A. Aucar, V. Bakken,
  K. G. Dyall, S. Dubillard, U. Ekstroem, E. Eliav, T. Enevoldsen, E.
  Fasshauer, T. Fleig, O. Fossgaard, L. Halbert, E. D. Hedegaard, T. Helgaker,
  J. Henriksson, M. Ilias, Ch. R. Jacob, S. Knecht, S. Komorovsky, O. Kullie,
  J. K. Laerdahl, C. V. Larsen, Y. S. Lee, H. S. Nataraj, M. K. Nayak, P.
  Norman, M. Olejniczak, J. Olsen, J. M. H. Olsen, Y. C. Park, J. K. Pedersen,
  M. Pernpointner, R. Di Remigio, K. Ruud, P. Salek, B. Schimmelpfennig, B.
  Senjean, A. Shee, J. Sikkema, A. J. Thorvaldsen, J. Thyssen, J. van Stralen,
  M. L. Vidal, S. Villaume, O. Visser, T. Winther, and S. Yamamoto (see
  http://diracprogram.org).}

\bibitem[{\citenamefont{Saue et~al.}(2020)\citenamefont{Saue, Bast, Gomes,
  Jensen, Visscher, Aucar, Di~Remigio, Dyall, Eliav, Fasshauer
  et~al.}}]{Saue:2020}
\bibinfo{author}{\bibfnamefont{T.}~\bibnamefont{Saue}},
  \bibinfo{author}{\bibfnamefont{R.}~\bibnamefont{Bast}},
  \bibinfo{author}{\bibfnamefont{A.~S.~P.} \bibnamefont{Gomes}},
  \bibinfo{author}{\bibfnamefont{H.~J.~A.} \bibnamefont{Jensen}},
  \bibinfo{author}{\bibfnamefont{L.}~\bibnamefont{Visscher}},
  \bibinfo{author}{\bibfnamefont{I.~A.} \bibnamefont{Aucar}},
  \bibinfo{author}{\bibfnamefont{R.}~\bibnamefont{Di~Remigio}},
  \bibinfo{author}{\bibfnamefont{K.~G.} \bibnamefont{Dyall}},
  \bibinfo{author}{\bibfnamefont{E.}~\bibnamefont{Eliav}},
  \bibinfo{author}{\bibfnamefont{E.}~\bibnamefont{Fasshauer}},
  \bibnamefont{et~al.}, \bibinfo{journal}{J.\ Chem.\ Phys.}
  \textbf{\bibinfo{volume}{152}}, \bibinfo{pages}{204104}
  (\bibinfo{year}{2020}).

\bibitem[{MRC()}]{MRCC2020}
\bibinfo{note}{M. K\'{a}llay, P. R. Nagy, D. Mester, Z. Rolik, G. Samu, J.
  Csontos, J. Cs\'{o}ka, P. B. Szab\'{o}, L. Gyevi-Nagy, B. H\'{e}gely, I.
  Ladj\'{a}nszki, L. Szegedy, B. Lad\'{o}czki, K. Petrov, M. Farkas, P. D.
  Mezei, and \'{a}. Ganyecz: The {\sc mrcc} program system: Accurate quantum
  chemistry from water to proteins, J. Chem. Phys. 152, 074107 (2020).”
  “{\sc mrcc}, a quantum chemical program suite written by M. K\'{a}llay, P.
  R. Nagy, D. Mester, Z. Rolik, G. Samu, J. Csontos, J. Cs\'{o}ka, P. B.
  Szab\'{o}, L. Gyevi-Nagy, B. H\'{e}gely, I. Ladj\'{a}nszki, L. Szegedy, B.
  Lad\'{o}czki, K. Petrov, M. Farkas, P. D. Mezei, and \'{a}. Ganyecz. See
  www.mrcc.hu.}

\bibitem[{\citenamefont{K\'{a}llay and Surj\'{a}n}(2001)}]{Kallay:1}
\bibinfo{author}{\bibfnamefont{M.}~\bibnamefont{K\'{a}llay}} \bibnamefont{and}
  \bibinfo{author}{\bibfnamefont{P.~R.} \bibnamefont{Surj\'{a}n}},
  \bibinfo{journal}{J.\ Chem.\ Phys.} \textbf{\bibinfo{volume}{115}},
  \bibinfo{pages}{2945} (\bibinfo{year}{2001}).

\bibitem[{\citenamefont{K\'{a}llay et~al.}(2002)\citenamefont{K\'{a}llay,
  Szalay, and Surj\'{a}n}}]{Kallay:2}
\bibinfo{author}{\bibfnamefont{M.}~\bibnamefont{K\'{a}llay}},
  \bibinfo{author}{\bibfnamefont{P.~G.} \bibnamefont{Szalay}},
  \bibnamefont{and} \bibinfo{author}{\bibfnamefont{P.~R.}
  \bibnamefont{Surj\'{a}n}}, \bibinfo{journal}{J.\ Chem.\ Phys.}
  \textbf{\bibinfo{volume}{117}}, \bibinfo{pages}{980} (\bibinfo{year}{2002}).

\bibitem[{\citenamefont{Oleynichenko et~al.}(2021)\citenamefont{Oleynichenko,
  Zaitsevskii, and Eliav}}]{EXPT_website}
\bibinfo{author}{\bibfnamefont{A.}~\bibnamefont{Oleynichenko}},
  \bibinfo{author}{\bibfnamefont{A.}~\bibnamefont{Zaitsevskii}},
  \bibnamefont{and} \bibinfo{author}{\bibfnamefont{E.}~\bibnamefont{Eliav}}
  (\bibinfo{year}{2021}), \bibinfo{note}{{EXP-T}, an extensible code for {F}ock
  space relativistic coupled cluster calculations (see
  \url{http://www.qchem.pnpi.spb.ru/expt})}.

\bibitem[{\citenamefont{Oleynichenko et~al.}(2020)\citenamefont{Oleynichenko,
  Zaitsevskii, and Eliav}}]{Oleynichenko_EXPT}
\bibinfo{author}{\bibfnamefont{A.~V.} \bibnamefont{Oleynichenko}},
  \bibinfo{author}{\bibfnamefont{A.}~\bibnamefont{Zaitsevskii}},
  \bibnamefont{and} \bibinfo{author}{\bibfnamefont{E.}~\bibnamefont{Eliav}}, in
  \emph{\bibinfo{booktitle}{Supercomputing}}, edited by
  \bibinfo{editor}{\bibfnamefont{V.}~\bibnamefont{Voevodin}} \bibnamefont{and}
  \bibinfo{editor}{\bibfnamefont{S.}~\bibnamefont{Sobolev}}
  (\bibinfo{publisher}{Springer International Publishing},
  \bibinfo{address}{Cham}, \bibinfo{year}{2020}), vol. \bibinfo{volume}{1331},
  pp. \bibinfo{pages}{375--386}.

\bibitem[{\citenamefont{Skripnikov}(2016)}]{Skripnikov:16b}
\bibinfo{author}{\bibfnamefont{L.~V.} \bibnamefont{Skripnikov}},
  \bibinfo{journal}{J.\ Chem.\ Phys.} \textbf{\bibinfo{volume}{145}},
  \bibinfo{pages}{214301} (\bibinfo{year}{2016}).

\bibitem[{\citenamefont{Prosnyak and Skripnikov}(2021)}]{Prosnyak:2021}
\bibinfo{author}{\bibfnamefont{S.~D.} \bibnamefont{Prosnyak}} \bibnamefont{and}
  \bibinfo{author}{\bibfnamefont{L.~V.} \bibnamefont{Skripnikov}},
  \bibinfo{journal}{Phys. Rev. C} \textbf{\bibinfo{volume}{103}},
  \bibinfo{pages}{034314} (\bibinfo{year}{2021}).

\bibitem[{\citenamefont{Skripnikov}(2020)}]{Skripnikov:2020e}
\bibinfo{author}{\bibfnamefont{L.~V.} \bibnamefont{Skripnikov}},
  \bibinfo{journal}{J.\ Chem.\ Phys.} \textbf{\bibinfo{volume}{153}},
  \bibinfo{pages}{114114} (\bibinfo{year}{2020}).

\bibitem[{\citenamefont{Prosnyak et~al.}(2020)\citenamefont{Prosnyak, Maison,
  and Skripnikov}}]{Prosnyak:2020}
\bibinfo{author}{\bibfnamefont{S.~D.} \bibnamefont{Prosnyak}},
  \bibinfo{author}{\bibfnamefont{D.~E.} \bibnamefont{Maison}},
  \bibnamefont{and} \bibinfo{author}{\bibfnamefont{L.~V.}
  \bibnamefont{Skripnikov}}, \bibinfo{journal}{J.\ Chem.\ Phys.}
  \textbf{\bibinfo{volume}{152}}, \bibinfo{pages}{044301}
  (\bibinfo{year}{2020}).

\bibitem[{\citenamefont{Skripnikov et~al.}(2017)\citenamefont{Skripnikov,
  Titov, and Flambaum}}]{Skripnikov:17b}
\bibinfo{author}{\bibfnamefont{L.~V.} \bibnamefont{Skripnikov}},
  \bibinfo{author}{\bibfnamefont{A.~V.} \bibnamefont{Titov}}, \bibnamefont{and}
  \bibinfo{author}{\bibfnamefont{V.~V.} \bibnamefont{Flambaum}},
  \bibinfo{journal}{Phys.\ Rev.\ A} \textbf{\bibinfo{volume}{95}},
  \bibinfo{pages}{022512} (\bibinfo{year}{2017}).

\bibitem[{\citenamefont{Dyall}(1998)}]{Dyall:98}
\bibinfo{author}{\bibfnamefont{K.~G.} \bibnamefont{Dyall}},
  \bibinfo{journal}{Theor. Chem. Acc.} \textbf{\bibinfo{volume}{99}},
  \bibinfo{pages}{366} (\bibinfo{year}{1998}).

\bibitem[{\citenamefont{Dyall}(2006)}]{Dyall:06}
\bibinfo{author}{\bibfnamefont{K.~G.} \bibnamefont{Dyall}},
  \bibinfo{journal}{Theor. Chem. Acc.} \textbf{\bibinfo{volume}{115}},
  \bibinfo{pages}{441} (\bibinfo{year}{2006}).

\bibitem[{\citenamefont{Dyall}(2012)}]{Dyall:12}
\bibinfo{author}{\bibfnamefont{K.~G.} \bibnamefont{Dyall}},
  \bibinfo{journal}{Theor. Chem. Acc.} \textbf{\bibinfo{volume}{131}},
  \bibinfo{pages}{1217} (\bibinfo{year}{2012}).

\bibitem[{\citenamefont{K\'{a}llay and Gauss}(2005)}]{Kallay:6}
\bibinfo{author}{\bibfnamefont{M.}~\bibnamefont{K\'{a}llay}} \bibnamefont{and}
  \bibinfo{author}{\bibfnamefont{J.}~\bibnamefont{Gauss}},
  \bibinfo{journal}{J.\ Chem.\ Phys.} \textbf{\bibinfo{volume}{123}},
  \bibinfo{eid}{214105} (pages~\bibinfo{numpages}{13}) (\bibinfo{year}{2005}).

\bibitem[{\citenamefont{Sikkema et~al.}(2009)\citenamefont{Sikkema, Visscher,
  Saue, and Ilia\v{s}}}]{Sikkema:2009}
\bibinfo{author}{\bibfnamefont{J.}~\bibnamefont{Sikkema}},
  \bibinfo{author}{\bibfnamefont{L.}~\bibnamefont{Visscher}},
  \bibinfo{author}{\bibfnamefont{T.}~\bibnamefont{Saue}}, \bibnamefont{and}
  \bibinfo{author}{\bibfnamefont{M.}~\bibnamefont{Ilia\v{s}}},
  \bibinfo{journal}{J.\ Chem.\ Phys.} \textbf{\bibinfo{volume}{131}},
  \bibinfo{pages}{124116} (\bibinfo{year}{2009}).

\bibitem[{\citenamefont{Tupitsyn et~al.}(1977--2002)\citenamefont{Tupitsyn,
  Deyneka, and Bratzev}}]{HFD}
\bibinfo{author}{\bibfnamefont{I.~I.} \bibnamefont{Tupitsyn}},
  \bibinfo{author}{\bibfnamefont{G.~B.} \bibnamefont{Deyneka}},
  \bibnamefont{and} \bibinfo{author}{\bibfnamefont{V.~F.}
  \bibnamefont{Bratzev}} (\bibinfo{year}{1977--2002}), \bibinfo{note}{{\sc
  hfd}, a program for atomic finite-difference four-component
  {D}irac-{H}artree-{F}ock calculations on the base of the HFD
  code~\cite{Bratzev:77}}.

\bibitem[{\citenamefont{Bratzev et~al.}(1977)\citenamefont{Bratzev, Deyneka,
  and Tupitsyn}}]{Bratzev:77}
\bibinfo{author}{\bibfnamefont{V.~F.} \bibnamefont{Bratzev}},
  \bibinfo{author}{\bibfnamefont{G.~B.} \bibnamefont{Deyneka}},
  \bibnamefont{and} \bibinfo{author}{\bibfnamefont{I.~I.}
  \bibnamefont{Tupitsyn}}, \bibinfo{journal}{Bull.\ Acad.\ Sci.\ USSR, Phys.\
  Ser.} \textbf{\bibinfo{volume}{41}}, \bibinfo{pages}{173}
  (\bibinfo{year}{1977}).

\bibitem[{\citenamefont{Tupitsyn}(2003)}]{HFDB}
\bibinfo{author}{\bibfnamefont{I.~I.} \bibnamefont{Tupitsyn}}
  (\bibinfo{year}{2003}), \bibinfo{note}{{\sc hfdb}, a program for atomic
  finite-difference four-component {D}irac-{H}artree-{F}ock-{B}reit
  calculations written on the base of the {\sc hfd} code~\cite{Bratzev:77}}.

\bibitem[{\citenamefont{Visser et~al.}(1987)\citenamefont{Visser, Aerts,
  Hegarty, and Nieuwpoort}}]{Visser:87}
\bibinfo{author}{\bibfnamefont{O.}~\bibnamefont{Visser}},
  \bibinfo{author}{\bibfnamefont{P.~J.~C.}~\bibnamefont{Aerts}},
  \bibinfo{author}{\bibfnamefont{D.}~\bibnamefont{Hegarty}}, \bibnamefont{and}
  \bibinfo{author}{\bibfnamefont{W.~C.}~\bibnamefont{Nieuwpoort}},
  \bibinfo{journal}{Chem.\ Phys.\ Lett.} \textbf{\bibinfo{volume}{134}},
  \bibinfo{pages}{34} (\bibinfo{year}{1987}).

\bibitem[{\citenamefont{Fricke}(1991)}]{Fricke18}
\bibinfo{author}{\bibfnamefont{B.}~\bibnamefont{Fricke}}
  (\bibinfo{year}{1991}), \bibinfo{note}{private communication cited in Ref.
  [18]}.

\bibitem[{\citenamefont{Fricke}(1984)}]{Fricke22}
\bibinfo{author}{\bibfnamefont{B.}~\bibnamefont{Fricke}}
  (\bibinfo{year}{1984}), \bibinfo{note}{private communication cited in Ref.
  [22]}.

\bibitem[{\citenamefont{Hartley and
  Martensson-Pendrill}(1991)}]{Martensson1991}
\bibinfo{author}{\bibfnamefont{A.~C.} \bibnamefont{Hartley}} \bibnamefont{and}
  \bibinfo{author}{\bibfnamefont{A.~M.} \bibnamefont{Martensson-Pendrill}},
  \bibinfo{journal}{J. Phys. B: At. Mol. Opt. Phys.}
  \textbf{\bibinfo{volume}{24}}, \bibinfo{pages}{1993} (\bibinfo{year}{1991}).

\bibitem[{\citenamefont{M\aa{}rtensson-Pendrill}(1995)}]{Martensson1995}
\bibinfo{author}{\bibfnamefont{A.-M.} \bibnamefont{M\aa{}rtensson-Pendrill}},
  \bibinfo{journal}{Phys. Rev. Lett.} \textbf{\bibinfo{volume}{74}},
  \bibinfo{pages}{2184} (\bibinfo{year}{1995}).

\bibitem[{\citenamefont{Zubova et~al.}(2014)\citenamefont{Zubova, Kozhedub,
  Shabaev, Tupitsyn, Volotka, Plunien, Brandau, and St\"ohlker}}]{Zubova:2014}
\bibinfo{author}{\bibfnamefont{N.~A.} \bibnamefont{Zubova}},
  \bibinfo{author}{\bibfnamefont{Y.~S.} \bibnamefont{Kozhedub}},
  \bibinfo{author}{\bibfnamefont{V.~M.} \bibnamefont{Shabaev}},
  \bibinfo{author}{\bibfnamefont{I.~I.} \bibnamefont{Tupitsyn}},
  \bibinfo{author}{\bibfnamefont{A.~V.} \bibnamefont{Volotka}},
  \bibinfo{author}{\bibfnamefont{G.}~\bibnamefont{Plunien}},
  \bibinfo{author}{\bibfnamefont{C.}~\bibnamefont{Brandau}}, \bibnamefont{and}
  \bibinfo{author}{\bibfnamefont{T.}~\bibnamefont{St\"ohlker}},
  \bibinfo{journal}{Phys. Rev. A} \textbf{\bibinfo{volume}{90}},
  \bibinfo{pages}{062512} (\bibinfo{year}{2014}).

\bibitem[{\citenamefont{Wansbeek et~al.}(2012)\citenamefont{Wansbeek,
  Schlesser, Sahoo, Dieperink, Onderwater, and Timmermans}}]{Wansbeek2012}
\bibinfo{author}{\bibfnamefont{L.~W.} \bibnamefont{Wansbeek}},
  \bibinfo{author}{\bibfnamefont{S.}~\bibnamefont{Schlesser}},
  \bibinfo{author}{\bibfnamefont{B.~K.} \bibnamefont{Sahoo}},
  \bibinfo{author}{\bibfnamefont{A.~E.~L.} \bibnamefont{Dieperink}},
  \bibinfo{author}{\bibfnamefont{C.~J.~G.} \bibnamefont{Onderwater}},
  \bibnamefont{and} \bibinfo{author}{\bibfnamefont{R.~G.~E.}
  \bibnamefont{Timmermans}}, \bibinfo{journal}{Phys. Rev. C}
  \textbf{\bibinfo{volume}{86}}, \bibinfo{pages}{015503}
  (\bibinfo{year}{2012}).

\bibitem[{\citenamefont{Wilson}(1988)}]{Wilson22}
\bibinfo{author}{\bibfnamefont{M.}~\bibnamefont{Wilson}}
  (\bibinfo{year}{1988}), \bibinfo{note}{private communication cited in Ref.
  [22]}.

\bibitem[{\citenamefont{Skripnikov}(2021)}]{Skripnikov:2021a}
\bibinfo{author}{\bibfnamefont{L.~V.} \bibnamefont{Skripnikov}},
  \bibinfo{journal}{J.\ Chem.\ Phys.} \textbf{\bibinfo{volume}{154}},
  \bibinfo{pages}{201101} (\bibinfo{year}{2021}).

\bibitem[{\citenamefont{Yerokhin}(2011)}]{Yerokhin:2011b}
\bibinfo{author}{\bibfnamefont{V.~A.} \bibnamefont{Yerokhin}},
  \bibinfo{journal}{Phys. Rev. A} \textbf{\bibinfo{volume}{83}},
  \bibinfo{pages}{012507} (\bibinfo{year}{2011}).

\bibitem[{\citenamefont{M{\"o}ller et~al.}(2016)\citenamefont{M{\"o}ller,
  Sierk, Ichikawa, and Sagawa}}]{Moller:2016}
\bibinfo{author}{\bibfnamefont{P.}~\bibnamefont{M{\"o}ller}},
  \bibinfo{author}{\bibfnamefont{A.~J.}~\bibnamefont{Sierk}},
  \bibinfo{author}{\bibfnamefont{T.}~\bibnamefont{Ichikawa}}, \bibnamefont{and}
  \bibinfo{author}{\bibfnamefont{H.}~\bibnamefont{Sagawa}},
  \bibinfo{journal}{At. Data Nucl. Data Tables}
  \textbf{\bibinfo{volume}{109-110}}, \bibinfo{pages}{1}
  (\bibinfo{year}{2016}).

\end{thebibliography}
\end{document}